\pdfoutput=1
\documentclass[twocolumn]{aastex701}
\usepackage{array,makecell}
\usepackage{multirow}

\shortauthors{Ng, Zhang, \& Chen}

\begin{document}

\title{Anomalously Strong Localized First Ionization Potential Effect Associated with a Solar Subflare}

\author[0000-0001-6206-0138]{Man-Hei Ng}
\affiliation{State Key Laboratory of Lunar and Planetary Sciences and CNSA Macau Center for Space Exploration and Science, Macau University of Science and Technology, Macau, People's Republic of China}
\email{marcosng.hei@gmail.com}

\author[0000-0002-4306-5213]{Xiaoping Zhang}
\affiliation{State Key Laboratory of Lunar and Planetary Sciences and CNSA Macau Center for Space Exploration and Science, Macau University of Science and Technology, Macau, People's Republic of China}
\email[show]{xpzhang@must.edu.mo}

\author[0000-0002-7289-642X]{P. F. Chen}
\affiliation{School of Astronomy and Space Science and Key Laboratory of Modern Astronomy and Astrophysics, Nanjing University, Nanjing, People's Republic of China}
\affiliation{State Key Laboratory of Lunar and Planetary Sciences and CNSA Macau Center for Space Exploration and Science, Macau University of Science and Technology, Macau, People's Republic of China}
\email{chenpf@nju.edu.cn}

\begin{abstract}
Plasma composition in the solar corona commonly differs from that of the photosphere, with the enhancement of low--first-ionization-potential (FIP) elements referred to as the FIP effect.
This phenomenon provides important diagnostics of energy and mass transport between different layers of the solar atmosphere.
In this work, we analyze an anomalously strong, localized FIP effect observed in active region~13486 associated with a subflaring episode on 2023 November 17, using multiwavelength observations combining high energy-resolution soft X-ray disk-integrated spectra obtained by the Macao Science Satellite-1B with spatially resolved EUV/UV and H$\alpha$ imaging from Hinode/EIS, SDO/AIA and HMI, and CHASE/HIS.
By investigating the temporal evolution of plasma composition in response to changes in magnetic field orientation, we provide new insight into the physical processes linking magnetic reconnection, ponderomotive force fractionation, and coronal abundance anomalies.
This work reveals that the anomalously strong enhancement of low-FIP elements is localized in regions with strongly inclined magnetic fields despite a subflare. %
We interpret these observations within the framework of the ponderomotive force fractionation model and propose that the inclined magnetic geometry enhances the transmission of upward-propagating magnetohydrodynamic waves by reducing reflection near the plasma-$\beta$$\simeq$1 layer, enhancing FIP fractionation associated with a consequential upward-directed ponderomotive force.
In addition, sustained chromospheric heating associated with chromospheric reconnection and flux cancellation appears to maintain the enhanced FIP effect for tens of minutes following the event. %
\end{abstract}


\section{Introduction} \label{sec_intro}

Understanding how the elemental composition of coronal plasma varies relative to the underlying photosphere is of great interest to the solar physics community. Such compositional variations provide key diagnostics for probing energy and mass transport between different layers of the solar atmosphere and for constraining the physical mechanisms responsible for coronal heating.
Early observations of the solar atmosphere in the 1960s revealed anomalous variations in elemental composition between the solar corona and the photosphere~\citep{pottaschLowerSolarCorona1963}. %
It was later established that these anomalies systematically depend on the first ionization potential (FIP) of the elements, i.e., low-FIP ($\lesssim$10~eV) elements, such as Ca, Mg, Fe, and Si, are typically enhanced in the corona by factors of 2--4 relative to their photospheric abundances.
In contrast, high-FIP elements (e.g., Ar, O, Ne) in the corona generally are relatively unaffected or similar to those in the photosphere~\citep[e.g.,][]{meyerBaselineCompositionSolar1985, meyerSolarstellarOuterAtmospheres1985, widing1989AbundanceVariationsOuter, widing1995AbundanceRatiosOxygen, feldmanElementalAbundancesUpper1992, feldmanElementAbundancesUpper2000, feldmanReviewFirstIonization2002, asplundChemicalCompositionSun2009, asplundChemicalMakeupSun2021, delzanna2014ElementalAbundancesTemperatures, dennisSOLARFLAREELEMENT2015}. %
This abundance variation, i.e., the enrichment of low-FIP elements in the corona relative to the underlying photosphere, is known as the FIP effect and is commonly quantified by the FIP bias, defined as the ratio of coronal to photospheric abundances for a given element measured on an absolute scale with respect to hydrogen.
However, measuring the absolute FIP bias (i.e., the abundance of each element relative to hydrogen) can be challenging with some instruments, e.g., extreme-ultraviolet (EUV) imaging spectroscopy. Thus, a relative FIP bias is often used to quantify the FIP effect, defined as the abundance ratio of a low-FIP element to a high-FIP element.
Such a relative measure, however, cannot unambiguously distinguish whether the observed FIP effect is caused by an enhancement of low-FIP elements, a depletion of high-FIP elements with respect to their photospheric abundances, or a combination of both processes. %
Moreover, in some situations the abundances of low-FIP elements  are depleted in the corona compared to the photosphere, a phenomenon referred to as the inverse FIP (iFIP) effect, i.e., $\text{FIP bias} < 1$.
The solar iFIP effect has been observed in localized regions of strong magnetic fields near sunspots within complex active regions (ARs; e.g.,~\citealp{doschekAnomalousRelativeAr2015, doschekMYSTERIOUSCASESOLAR2016, doschekSunspotsStarspotsElemental2017, brooksDiagnosticCoronalElemental2018, bakerTransientInverseFIPPlasma2019, bakerCanSubphotosphericMagnetic2020, bakerSearchingEvidenceSubchromospheric2024}), as well as during the peak phases of flare eruptions (e.g.,~\citealp{warrenABSOLUTECALIBRATIONEUV2014, katsudaInverseFirstIonization2020, ngUnveilingMassTransfer2024}).

A widely accepted theoretical framework for explaining both the FIP and iFIP effects is the ponderomotive force fractionation model~\citep{lamingUnifiedPictureFirst2004, lamingNONWKBMODELSFIRST2009, lamingNONWKBMODELSFIRST2012, lamingFirstIonizationPotential2017, lamingFIPInverseFIP2015, lamingFIPInverseFIPEffects2021, lamingElementAbundancesNew2019, lamingElementAbundancesImpulsive2023}.
This model proposed that FIP/iFIP fractionation is driven by the ponderomotive force associated with magnetohydrodynamic (MHD) waves in the solar chromosphere, acting as an agent of ion--neutral separation.
The ponderomotive force is caused by wave reflection and/or refraction in regions of steep density and magnetic-field gradients and acts exclusively on charged particles.
The wave propagation direction determines the direction of the ponderomotive force acting on ions. %
In the chromosphere, low-FIP elements are more easily ionized than high-FIP elements, and therefore low-FIP ions are preferentially affected by the ponderomotive force and separated from high-FIP neutrals.
As a result, these ions are either accelerated upward into the corona (leading to coronal enrichment) or driven downward toward the photosphere or lower chromosphere (leading to coronal depletion), thereby giving rise to the observed FIP or iFIP effects, respectively.
Recent observations by \citet{murabito2024ObservationAlfvenWave} provided clear evidence for the reflection of coronal Alfv\'en waves in the chromosphere, establishing a critical observational link between coronal abundance anomalies and the theoretical framework of the ponderomotive force fractionation model.
Their work also indicated that higher frequency waves propagate deeper into the chromosphere before reflecting back upward, and the differences in how resonant and nonresonant waves interact with coronal magnetic loops suggest that elemental fractionation is a nonlocal process, dependent on the degree to which the wave is resonant with the loop.
Here, resonance refers to the condition where the Alfv\'en wave travel time between the two loop footpoints is an integral number of half wave periods.
In addition, \citet{mihailescuIntriguingPlasmaComposition2023} further showed that the height at which fractionation occurs depends on the wave mode.
Resonant waves of coronal origin drive fractionation predominantly near the top of the chromosphere, because the wave energy reflects around the upper chromosphere due to steep density variation and remains concentrated within the coronal cavity rather than at chromospheric footpoints. %
In contrast, nonresonant waves (particular at higher frequencies), originating either in the corona or the photosphere, propagate deeper into the chromosphere, where they deposit their energy and drive stronger fractionation at lower chromospheric heights than that produced by resonant modes.

In contrast to the case of closed magnetic loops described above, open field regions such as coronal holes typically exhibit much weaker fractionation, with FIP bias values generally ranging from 1 to 2~\citep{feldman1998CoronalCompositionSolar, brooks2011ESTABLISHINGCONNECTIONACTIVE, bakerPlasmaCompositionSigmoidal2013, lamingFirstIonizationPotential2017, leeCoronalAbundanceFractionation2025}.
In such magnetic configurations, waves easily escape into the solar wind and cannot reflect multiple times at the chromospheric footpoints as in closed loop case.
These waves have little to no resonance in the coronal cross sections of the open field structures, wave energy is not efficiently trapped, resulting in little or no elemental fractionation~\citep[][]{bakerPlasmaCompositionSigmoidal2013, bakerAlfvenicPerturbationsSunspot2021, lamingFIPInverseFIP2015, lamingFirstIonizationPotential2017, leeCoronalAbundanceFractionation2025}.
We note in passing that measurements of solar wind abundances indicate intermediate-FIP elements (e.g., elements S, P, and C; which are technically high-FIP elements with the lowest FIPs) are enhanced in corotating interaction regions relative to those in gradual solar energetic particles and the closed-loop solar corona~\citep{reamesAbundancesIonizationStates2018, lamingElementAbundancesNew2019, kurodaMagneticFieldGeometry2020}.
These subtle variations in FIP fractionation have been further explored by \citet{lamingElementAbundancesImpulsive2023}, who suggested that in open-field configurations lacking resonant Alfv\'en waves, fractionation of intermediate-FIP elements can occur if the magnetic field strength is sufficiently high.
Their model showed that the ponderomotive force acts throughout much of the chromosphere, extending from the top down to just above the plasma-$\beta$=1.2 equipartition layer, where the sound speed and Alfv\'en speed become comparable.
A stronger magnetic field pushes this layer to lower altitudes, where the background hydrogen is predominantly neutral while helium remains undepleted.
Under such conditions, sulfur behaves more similarly to other low-FIP elements, so can the fractionation of S develop.
That said, in the general sense, elemental fractionation should be more evident in flaring loops, since various types of MHD waves are generated and subsequently trapped, bouncing back and forth within the loop structure.
Indeed, the compositional changes during solar flares have been extensively investigated over the past decade~\citep[e.g.,][]{dennisSOLARFLAREELEMENT2015, mondal2021EvolutionElementalAbundances, mithun2022SoftXRaySpectral, suarezEstimationsElementalAbundances2023, telikicherlaInvestigatingSoftXRay2024, ngUnveilingMassTransfer2024, toSpatiallyResolvedPlasma2024, ngPeculiarFeatureFirst2025}, indicating that chromospheric evaporation contributes significantly to the redistribution of elemental abundances in the corona during flares, with subsequent elemental fractionation potentially caused by reconnection-driven Alfv\'en waves in flaring loops. %
Beyond these effects, flaring loops differ from coronal loops not only in the intensity of MHD waves but also in their much more dynamic change of the magnetic configuration, since coronal loops are more or less stable, but flaring loops are contracting during energy release~\citep{forbesReconnectionFieldLine1996, chen1999FlaringLoopMotion}.
However, only a few studies~\citep[e.g.,][]{stangaliniSpectropolarimetricFluctuationsSunspot2021, bakerAlfvenicPerturbationsSunspot2021, murabitoInvestigatingOriginMagnetic2021} have made attempts to investigate the role of magnetic change in elemental fractionation.
Despite the highly dynamic nature of the flaring loops, the role of evolving magnetic field geometry in regulating coronal plasma composition remains relatively unexplored.

In this work, we investigate an anomalously strong, localized FIP effect observed in a subflare located at AR~13486 on 2023 November 17 using high energy-resolution soft X-ray (SXR) spectroscopy and spatially resolved EUV/UV and H$\alpha$ imaging, and explore the physical mechanisms responsible for the observed coronal abundance anomalies. %
Section~\ref{sec_obs} outlines the observations of AR~13486 and details the multiwavelength data analysis, including the wavelet analysis.
The results are presented in Section~\ref{sec_results}, which are followed by a discussion of the observed fractionation in Section~\ref{sec_discussion} and a summary in Section~\ref{sec_summary}.

\section{Observations and Data Analysis} \label{sec_obs}

\subsection{Overview of AR~13486} \label{sec_event}

\begin{figure*}[ht!]%
\plotone{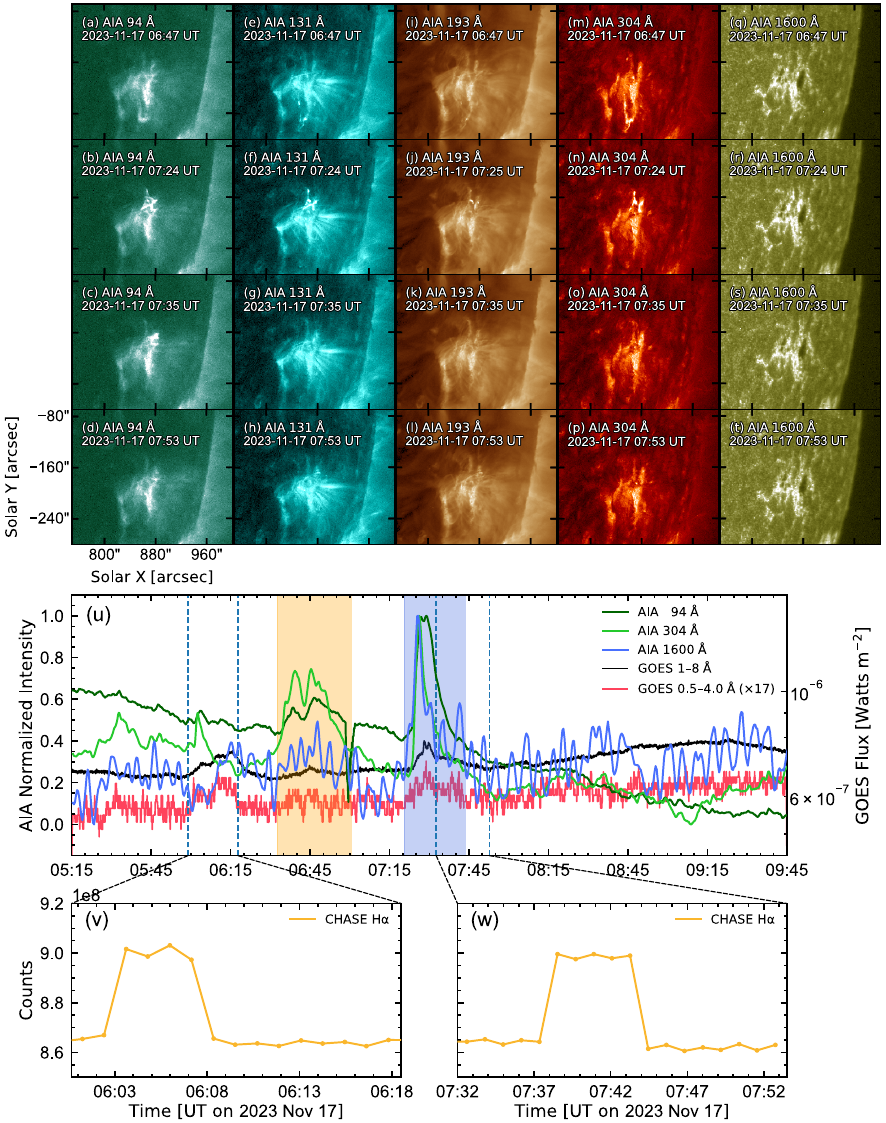}
\caption{Overview of the targeted AR~13486 observed on 2023 November 17. %
Context images from SDO/AIA, with the field of view (FOV) spanning from the bottom-left corner at [750\arcsec, \textminus280\arcsec] to the top-right corner at [1000\arcsec, \textminus70\arcsec], are displayed in the upper panels~(a)--(t).
Each row corresponds to a representative time during the evolution of the micro and subflaring activities associated with the AR, while each column shows observations in different AIA channels: (a)--(d) 94~{\AA}, (e)--(h) 131~{\AA}, (i)--(l) 193~{\AA}, (m)--(p) 304~{\AA}, and (q)--(t) 1600~{\AA}. %
The channels and selected observation times are indicated in the legends.
The following horizontal panel~(u) shows the normalized light curves from the defined FOV for the AIA 94~{\AA} (dark green), 304~{\AA} (bright green), and 1600~{\AA} (blue), with the soft X-ray (SXR) flux from GOES 1--8~{\AA} (black) and 0.5--4.0~{\AA} (red) overlaid. %
The AIA curves are smoothed using a moving average with a boxcar width of 5, and the GOES curves are smoothed using a median filter.
Orange and blue overlays highlight the time intervals corresponding to microflare (peaking at 06:45~UT in the 304~{\AA} channel) and subflare (peaking at 07:26~UT in the 304~{\AA} channel) activity, respectively.
Dashed blue vertical lines in the middle panel indicate the time intervals covered by CHASE H$\alpha$ line-core observations, with the corresponding light curves from the same FOV plotted in the bottom panels: (v) 05:58--06:18~UT and (w) 07:32--07:53~UT.
}{}
\label{fig_event}
\end{figure*}

The active region examined in this study initially formed in the southern hemisphere of the Sun on 2023 November 12 as newly emerging magnetic flux. It was designated NOAA AR~13486 on November 13 at S09W20 and rotated out of view on November 19. Throughout its passage across the visible disk, the region produced no major flares and evolved into a bipolar sunspot group, categorized as a $\beta$ configuration in the Mount Wilson (Hale) classification. %
On November 17, however, we observed that the region produced several small-scale flaring episodes, including several subflares and microflares near the west limb (S08W75). %
Figures~\ref{fig_event}(a)--(t) show observations from the Atmospheric Imaging Assembly~\citep[AIA;][]{lemenAtmosphericImagingAssembly2012} onboard the Solar Dynamics Observatory~\citep[SDO;][]{pesnellSolarDynamicsObservatory2012} on 2023 November 17, covering a field of view (FOV) from [750\arcsec, \textminus280\arcsec] to [1000\arcsec, \textminus70\arcsec].
The panels are arranged such that each column corresponds to one of the five AIA channels (94~{\AA}, 131~{\AA}, 193~{\AA}, 304~{\AA}, and 1600~{\AA}), while each row corresponds to one of four representative times illustrating the evolution of the small-scale flaring activity within AR~13486, which are (1) 06:47~UT, capturing a microflare prior to the subflare; (2) 07:24~UT, during the main phase of the subflare; (3) 07:35~UT, during the decay phase of the subflare; and (4) 07:53~UT, coinciding with the start of the EIS raster (see Section~\ref{sec_eisFIP}).  %

As several other active regions (ARs 13486--13489) were present on the solar disk during this interval, it is necessary to verify that AR~13486 was the principal source region contributing to the SXR emission that dominates the Sun-as-a-Star measurements by GOES and Macao Science Satellite-1B~\citep[MSS-1B;][]{zhangNovelGeomagneticSatellite2023, kongPrefaceSpecialIssue2023} presented later in this work (Section~\ref{sec_mssFIP}).
To establish this, we carried out a quantitative comparison between the GOES 1--8~{\AA} SXR flux and AIA light curves, specifically the 94~{\AA} (dark green), 304~{\AA} (bright green), and 1600~{\AA} (blue) channels, extracted from all visible active regions on the disk.
We note that \citet{ngCalibrationSolarXray2025} demonstrated quantitative agreement between the GOES and MSS-1B SXR flux measurements.
As shown by the black curve in Figure~\ref{fig_event}(u), the GOES 1--8~{\AA} profile exhibits a subtle enhancement between 07:20 and 07:43~UT relative to the otherwise flat background level (blue-shaded interval). %
The subflaring episode began at approximately 07:27~UT, showed two closely spaced peaks at 07:28 and 07:29~UT, and decayed back to the background level by about 07:50~UT, after which the overall background emission gradually increased. The GOES 0.5--4.0~{\AA} channel is plotted as the overlaid red curve. %
Among all ARs examined, the temporal variations of AR~13486 light curve (extracted from the FOV defined earlier and shown in Figure~\ref{fig_event}) exhibited the closest correspondence with the GOES SXR flux, indicating that AR~13486 was the dominant source region responsible for the observed subflaring emission.
In addition, we detected the earlier microflaring episode within the same FOV, occurring between 06:32 and 07:00~UT and highlighted it by the orange-shaded interval in Figure~\ref{fig_event}(u), with its corresponding context image shown in the first row of Figure~\ref{fig_event}.
Furthermore, H$\alpha$ line-core light curves (integrated over a Gaussian profile within $\pm$1~{\AA} of 6562.8~{\AA}) observed by the H$\alpha$ Imaging Spectrograph~\citep[HIS;][]{liuTechnologiesHaImaging2022, qiuCalibrationProceduresCHASE2022} onboard the Chinese H$\alpha$ Solar Explorer~\cite[CHASE;][]{liChineseHaSolar2019, liChineseHaSolar2022} are extracted from the same FOV are plotted in panels~(v) and (w) for the 05:58--06:18~UT and 07:32--07:53~UT intervals, respectively. The associated time intervals are also marked by pairs of dashed blue vertical lines in panel~(u).
These H$\alpha$ data are used in part of the wavelet analysis discussed in Sections~\ref{sec_wavelet} and \ref{sec_inclined}.

\subsection{MSS-1B SXR Observations and Spectral Analysis}

Observations of the Sun-as-a-Star were acquired by the Solar X-ray Detector~\citep[SXD;][]{shiDesignSolarXray2023} onboard the Macao Science Satellite-1B.
This instrument provides solar SXR observations through one of its primary components, i.e., the Soft X-ray Detection Units~(SXDUs), which comprise two detectors with aperture areas of 0.17~cm$^2$ (SXDU1) and 0.005~cm$^2$ (SXDU2), respectively, and a common field of view of $\pm2^{\circ}$.
The SXDUs deliver disk-integrated SXR and a little hard X-ray spectra in the energy range of 0.7--24~keV sampled across 2048 spectral channels, with a high energy resolution of 0.14~keV at 5.9~keV and a temporal cadence of 1~s.
The instrument specifications, detector design, and data calibration of the SXDUs are described in detail by \citet{shiDesignSolarXray2023}, \citet{zuoDesignTechnologyReview2024}, and \citet{ngCalibrationSolarXray2025}.
The data were further processed into a format compatible with the Object Spectral Executive~\citep[OSPEX, an X-ray spectral analysis package;][]{tolbertOSPEXObjectSpectral2020}, and the corresponding detector response matrix (DRM), which includes the instrumental effects, was produced for generating the theoretical model spectrum used in the present analysis, as detailed in \citet{ngPeculiarFeatureFirst2025}.
In this study, we analyzed the disk-integrated SXR spectra obtained on 2023 November 17, during which several active regions (ARs 13486--13489) were present on the solar disk, with subflaring activity in AR~13486 dominating the observed emission (see Section~\ref{sec_event}). For the analysis of the AR subflaring activity, we used the spectra recorded by the larger-aperture soft X-ray detector (SXDU1), which provide better counting statistics for the measurements. %

The solar SXR spectrum is characterized by a steeply decreasing continuum resulting from free--free, free-bound, and two-photon radiative processes, as well as various emission lines from different ionization states of elements such as Ca, Fe, S, and Ar.
These spectral components were modeled using the \texttt{CHIANTI} atomic database~\citep{dereCHIANTIAtomicDatabase1997} version 10.1~\citep{dere2023CHIANTIAtomicDatabase}, enabling the derivation of the plasma temperature ($T$), emission measure (EM, defined as $N_e^2 V$, where $N_e$ is the electron density and $V$ the emitting volume), and a set of elemental abundances.
For the spectral fitting, we adopted the same method as in \citet{ngPeculiarFeatureFirst2025}, which is summarized as follows.
We utilized the two-temperature (2-$T$) model \texttt{2vth\_abun} available in OSPEX. This model generates the theoretical spectrum as the sum of two isothermal components that share a common set of elemental abundances. The two components represent a cooler component associated with the background quiescent Sun and a hotter component associated with the subflaring activity within the AR. %
The best-fit parameters ($T$, EM, and abundances) were obtained through the iterative fitting procedure in OSPEX, which minimizes the reduced chi-squared statistic used to measure the goodness of fit between the observed and model-predicted count rates of the spectrum. %
The reduced chi-squared ($\chi_r^2$) value is defined as the sum, over all energy bins included in the fitting range, of the squared difference between the observed and model-predicted count rates divided by the model-predicted variance (assuming Poisson statistics), and then normalized by the number of degrees of freedom. %
The absolute abundances (i.e., relative to hydrogen) for the corona and photosphere are adopted from \citet{feldmanElementalAbundancesUpper1992} and \citet{asplundChemicalCompositionSun2009}, respectively. The ionization fractions are taken from \citet{dere2023CHIANTIAtomicDatabase}.
Furthermore, the best-fit set of elemental abundances is expressed in terms of FIP bias to quantify the FIP effect, with each FIP bias defined as the ratio of the measured coronal abundance of an element to its photospheric abundance.

Here, we aim to investigate the evolution of elemental composition before and after the AR subflaring activity, and how it is associated with the wave oscillations (Figure~\ref{fig_event}(u)) and magnetic configuration. We fitted the spectra for the following time intervals: (a) 03:20--03:45~UT (quiet corona), (b) 07:27--07:33~UT (during the main phase of the subflare), and (c) 07:33--08:05~UT (covering the decay to post-decay phase of the subflare) on 2023 November 17.
Since the emission produced by the subflaring activity and the quiet corona is much weaker than that of a typical flare, we selected a reasonable size of time bin for each analyzed spectrum to ensure sufficient counting statistics. In addition, the spectra were rebinned using a bin size of 7, corresponding to 82.7~keV per bin, to further improve the counting statistics.
To retain time-resolved diagnostic feasibility, we extended the preflare interval up to 4~hr before the event to represent the quiet corona within the available data, which allows us to track the temporal evolution of the elemental composition.

The observed spectra, shown in Figure~\ref{fig_mssSpec20231117}, clearly show several resolved emission lines, including Ca at 3.9~keV, S at 2.5~keV and Ar at 3.2~keV. The spectral fitting was performed over the energy range of 2.2--8.5~keV.
It is noted that, due to limited counting statistics at higher energies, the Fe line complex at $\sim$6.7~keV is either absent or too indistinct to allow a reliable determination of the Fe abundance. The results of Fe abundance are therefore not reported here.
Nonetheless, we verified that the derived temperatures, emission measures, and abundances of the other elements remain unaffected regardless whether the Fe abundance parameter is fixed or allowed to vary. %
Since the derived temperature is generally governed by the slope of the continuum, we retain the upper bound of the fitting range at 8.5~keV to ensure a reliable determination of the hotter component.
Section~\ref{sec_mssFIP} presents the temporal evolution of the temperatures, emission measures, and FIP biases of Ca, S, and Ar, associated with the AR subflaring activity, as derived from MSS-1B/SXDU1.

\subsection{EIS EUV Observations and MCMC-DEM Method}

We leveraged the spatially-resolved capabilities of the Extreme-ultraviolet Imaging Spectrometer~\citep[EIS;][]{culhaneEUVImagingSpectrometer2007} on board the Hinode spacecraft~\citep{kosugiHinodeSolarBMission2007} to analyze elemental composition.
Hinode/EIS records extreme ultraviolet (EUV) spectra through spatial raster scanning, covering two wavelength bands: 171--212~{\AA} (short-wavelength) and 245--291~{\AA} (long-wavelength), with a spectral resolution of approximately 22~m{\AA} and a spatial plate scale of 1\arcsec/pixel. %

The Hinode/EIS spectroscopic observation used in this study was taken on 2023 November 17, from 07:53 to 08:54~UT, targeting at AR~13486 and capturing the region during the post-decay phase of the subflare activity.
Key details of the EIS raster scan are summarized in Table~\ref{tab_EIS_studies}. %
We obtained the level-1 HDF5 data using the EIS Python Analysis Code~\citep[\texttt{EISPAC}\footnote{\href{https://eispac.readthedocs.io}{https://eispac.readthedocs.io}};][]{webergEISPACEISPython2023}.
The data were preprocessed using the \texttt{eis\_prep.pro} routine within Solar SoftWare~\citep[SSW;][]{freelandDataAnalysisSolarSoft1998}, which performs camera dark current and pedestal subtraction, cosmic ray removal, correction of hot, warm, and dusty pixels, and radiometric calibration.
Additional corrections were applied to account for instrumental effects such as orbital variation, detector offsets between different spectral windows, and slit tilt.
For this analysis, we adopted the absolute calibration presented by \citet{warrenABSOLUTECALIBRATIONEUV2014}, which compensates for sensitivity degradation over time and wavelength, and assumed a 23\% uncertainty in the line intensities~\citep{langLaboratoryCalibrationExtremeUltraviolet2006}. %

To diagnose the elemental composition, we followed the well-established method described in \citet{brooks2011ESTABLISHINGCONNECTIONACTIVE} and \citet{brooksFullSunObservationsIdentifying2015}, quantifying FIP bias values using intensity ratios of spectral line pairs between low- and high-FIP elements from the EIS raster scan: \ion{Ca}{14} 193.87~{\AA}/\ion{Ar}{14} 194.40~{\AA}, \ion{Fe}{16} 262.98~{\AA}/\ion{S}{13} 256.69~{\AA}, and \ion{Si}{10} 258.38~{\AA}/\ion{S}{10} 264.23~{\AA}.
All atomic data were obtained from the \texttt{CHIANTI} atomic database~\citep{dereCHIANTIAtomicDatabase1997}, version 10.1~\citep{dere2023CHIANTIAtomicDatabase}, which was used to compute line intensity ratios as a function of electron density and to calculate line contribution functions $G(n,T)$, adopting the photospheric abundances of \citet{scottElementalCompositionSun2015b, scottElementalCompositionSun2015a} and the ionization equilibrium from \citet{dere2023CHIANTIAtomicDatabase}. %
The ionization equilibrium peak temperatures for the corresponding ions are $\log\,(T_\mathrm{max}~[\mathrm{K}])=6.55$ for \ion{Ca}{14}, $6.55$ for \ion{Ar}{14}, $6.45$ for \ion{Fe}{16}, $6.40$ for \ion{S}{13}, $6.15$ for \ion{Si}{10}, and $6.15$ for \ion{S}{10}, respectively.
The electron density was subsequently inferred from the density-sensitive diagnostic line ratio of \ion{Fe}{13} 202.04~\AA/\ion{Fe}{13} 203.83~{\AA}.
To reduce the influence of temperature and density effects on the FIP bias diagnostics, we reconstructed the differential emission measure (DEM, the amount of plasma per unit temperature interval as a function of temperature) using a series of Fe emission lines from different ionization stages (see Table~\ref{tab_EIS_studies}), where we assumed the previously derived \ion{Fe}{13} density. %
The DEM inversion was carried out using the Markov Chain Monte Carlo (MCMC) algorithm~\citep{kashyapMarkovChainMonteCarlo1998}, with sampling implemented via the \texttt{emcee} package\footnote{\href{https://emcee.readthedocs.io}{https://emcee.readthedocs.io}}~\citep{foreman-mackeyEmceeMCMCHammer2013}, through which we performed 500 MCMC calculations to minimize the differences between the observed and theoretical line intensities and obtain the best-fit solution for the DEM from the posterior probability distribution of the model parameters. %

As mentioned, the best-fit DEM was constructed from the low-FIP Fe lines (most of which lie in the EIS short wavelength band).
Since the emission intensities of low-FIP ions, including \ion{Ca}{14} 193.87~{\AA}, \ion{Fe}{16} 262.98~{\AA}, and \ion{Si}{10} 258.38~{\AA}, are expected to be similarly enhanced in the corona due to the FIP effect, we scaled the Fe-based DEM using their observed-to-predicted intensity ratios, yielding the low-FIP--scaled DEM for each element used to evaluate the FIP bias factors.
Note that the Si and S lines are located in the EIS long wavelength band and are therefore related to the cross-detector calibration problem discussed in \citet{brooksFullSunObservationsIdentifying2015}. %
The predicted intensities of the high-FIP lines \ion{Ar}{14} 194.40~{\AA}, \ion{S}{13} 256.69~{\AA}, and \ion{S}{10} 264.23~{\AA}, which were observed within the same EIS wavelength bands as the corresponding paired low-FIP lines and, unlike the low-FIP lines, are assumed to have photospheric abundances, can nonetheless be reliably computed using the same low-FIP--scaled DEM and compared with their observed intensities to derive the FIP bias factors for Ca/Ar, Fe/S, and Si/S, respectively (as detailed in \citealp{brooksFullSunObservationsIdentifying2015}). %
For reference, all emission lines used in this analysis are summarized in Table~\ref{tab_EIS_studies}, and that all FIP bias diagnostics reported in this work were obtained employing the MCMC-based DEM method described above (see footnote\footnote{The MCMC-based DEM method used in this work is implemented in the \texttt{demcmc\_FIP} package, which is built on the \texttt{demcmc} package~(\href{https://demcmc.readthedocs.io}{https://demcmc.readthedocs.io}) and is available at \href{https://github.com/andyto1234/demcmc_FIP}{https://github.com/andyto1234/demcmc\_FIP}~(v0.5; Git commit \texttt{078378e}).}), which, as validated by \citet{brooksFullSunObservationsIdentifying2015}, yields FIP bias factors with an uncertainty of approximately 30\%, provided that the uncertainty in the EIS line intensities remains below 40--50\%.
 
\begin{deluxetable}{ll}[t]
\tabletypesize{\footnotesize}
\centering
\tablecolumns{2}
\tablecaption{EIS Study Details and Emission Lines Considered. \label{tab_EIS_studies}}
\tablehead{ \colhead{Study Details} & \colhead{Specification} } %
\startdata
Study Acronym & HPW021VEL260x512v2\\
Study Number & 569 \\
Raster ID & 539 \\
Raster Observation Time & 17/11/2023 07:53--08:54~UT\\
Rastering & 2\arcsec~slit, 87 positions, 3\arcsec~coarse step\\
Raster FOV & 260\arcsec $\times$ 512\arcsec\\ %
Exposure Time & 40~s\\
Total Raster Time & $1^\mathrm{h}0^\mathrm{m}45^\mathrm{s}$\\
Reference Spectral Window & \ion{Fe}{12} 195.12~\AA \\
Density Diagnostic Lines & \ion{Fe}{13} 202.04~\AA/\ion{Fe}{13} 203.83~\AA \\
\makecell[l]{Composition Diagnostic Lines\\~\\~\\} & \makecell[l]{
\ion{Ca}{14} 193.87~\AA/\ion{Ar}{14} 194.40~\AA,\\
\ion{Fe}{16} 262.98~\AA/\ion{S}{13} 256.69~\AA,\\
\ion{Si}{10} 258.38~\AA/\ion{S}{10} 264.23~\AA}\\
\makecell[l]{Emission Lines for DEM\\~\\~\\~\\~\\~\\~\\~\\~\\~\\} & \makecell[l]{
\ion{Fe}{8} 185.213~\AA,
\ion{Fe}{8} 186.601~\AA,\\
\ion{Fe}{9} 188.497~\AA,
\ion{Fe}{9} 197.862~\AA,\\
\ion{Fe}{10} 184.536~\AA,
\ion{Fe}{11} 188.216~\AA,\\
\ion{Fe}{11} 188.299~\AA,
\ion{Fe}{12} 186.880~\AA,\\
\ion{Fe}{12} 192.394~\AA,
\ion{Fe}{12} 195.119~\AA,\\
\ion{Fe}{13} 202.044~\AA,
\ion{Fe}{13} 203.826~\AA,\\
\ion{Fe}{14} 264.787~\AA,
\ion{Fe}{14} 270.519~\AA,\\
\ion{Fe}{15} 284.160~\AA,
\ion{Fe}{16} 262.984~\AA,\\
\ion{Fe}{17} 254.870~\AA,
\ion{Fe}{23} 263.760~\AA,\\
\ion{Fe}{24} 255.100~\AA
}\\
\enddata
\end{deluxetable}

\subsection{Wavelet Analysis} %

One can readily observe from Figure~\ref{fig_event} that both the AIA 1600~{\AA} and CHASE H$\alpha$ line-core light curves exhibit clear signatures of oscillations, motivating a quantitative investigation of their dominant periods and temporal evolution.
To this end, we applied a Morlet wavelet analysis to the light-curve time series from AIA and CHASE, following the approach described in \citet{torrencePracticalGuideWavelet1998}.\footnote{\href{https://atoc.colorado.edu/research/wavelets}{https://atoc.colorado.edu/research/wavelets}.}
Periodicity in the time series was assessed at the 95\% global confidence level, determined using the method of \citet{auchereFOURIERWAVELETANALYSIS2016}, which accounts for the total number of degrees of freedom in the wavelet power spectra.
In accordance with this method, both the AIA and CHASE time series were apodized with a Hann window and transformed into Fourier power spectra to model the background arising from stochastic, frequency-dependent fluctuations. %
The background models fitted to both spectra follow the form $\sigma(\nu) = A\nu^s + C$, comprising a power-law component and a constant offset, where $\sigma(\nu)$ represents the mean power expected at frequency $\nu$ and $s$ is the power-law index.
These fitted backgrounds were then used to compute the 95\% local and global confidence levels, against which the time-averaged wavelet power spectra were compared to identify significant periodicities.
The wavelet results are presented in Section~\ref{sec_wavelet}. %

\section{Results} \label{sec_results}

\subsection{Sun-as-a-Star FIP Bias Measurements} \label{sec_mssFIP}

\begin{figure*}[ht!]%
\centering
\plotone{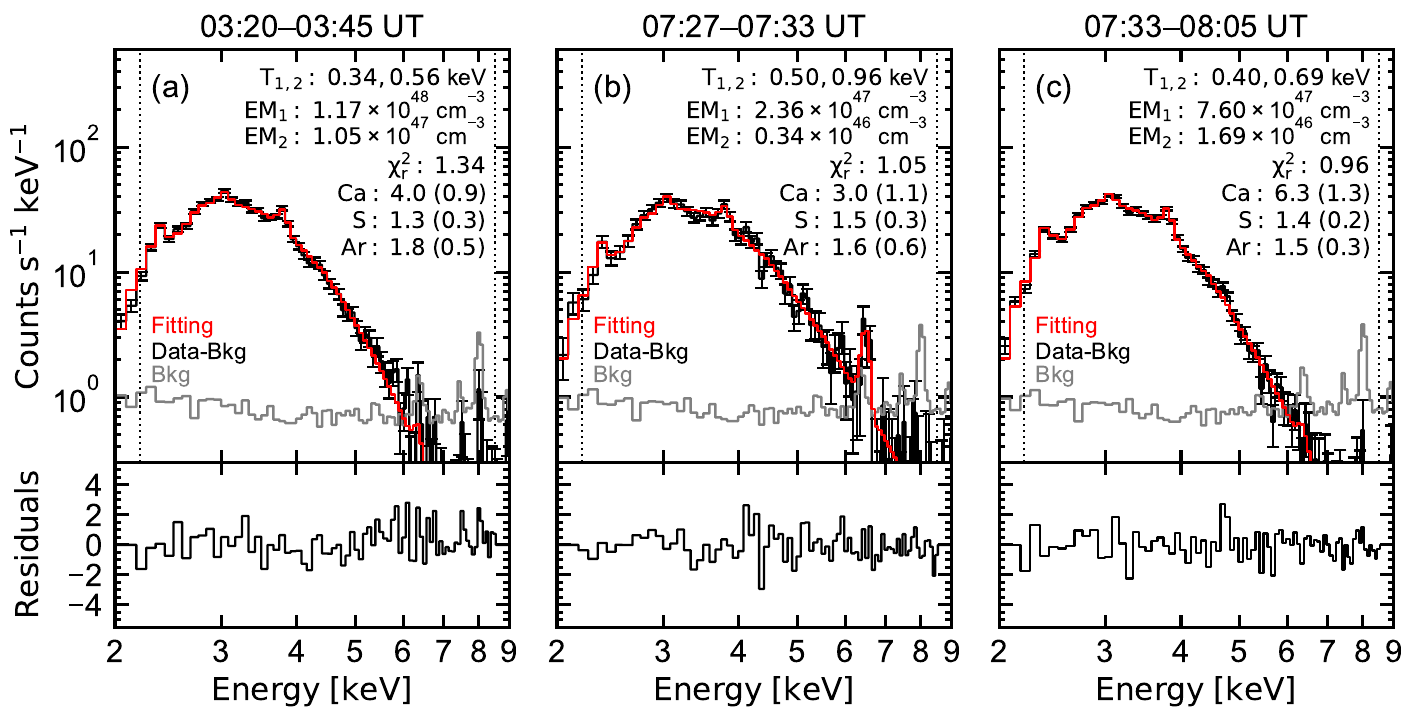}
\caption{Fitted MSS-1B SXR spectra on 2023 November 17 for (a) 03:20--03:45~UT, (b) 07:27--07:33~UT, and (c) 07:33--08:05~UT, obtained using the 2-$T$ model (\texttt{2vth\_abun}).
The selected fitting time periods are described in the text (see Section~\ref{sec_mssFIP}).
In each panel, the upper row displays the model fit (red curve), background-subtracted spectrum (black histogram), and background spectrum (gray histogram), all in units of $\mathrm{counts~s^{-1}~keV^{-1}}$.
The best-fit temperatures and emission measures, together with the FIP biases for Ca, S, and Ar (with 1$\sigma$ uncertainties), are indicated in the upper-right corner of each panel, along with the reduced chi-square ($\mathrm{\chi_r^2}$) values.
The FIP bias is defined as the ratio of the measured elemental abundance to its photospheric one, with coronal abundances adopted from \citet{feldmanElementalAbundancesUpper1992} and photospheric abundances from \citet{asplundChemicalCompositionSun2009}.
The lower row shows the normalized residuals, defined as the differences between the observed and best-fit fluxes divided by the $1\sigma$ statistical uncertainties.
}
\label{fig_mssSpec20231117}
\end{figure*}

Figure~\ref{fig_mssSpec20231117} shows the spectral fits for the three selected intervals, derived using a two-temperature (2-$T$) model. For each panel, the legend annotates the best-fit parameters, including the two temperatures ($T$), two emission measures (EMs), the elemental abundances of Ca, S, and Ar (with $\pm 1\sigma$ uncertainties), and the reduced chi-square ($\chi_r^2$) value used to evaluate the goodness of fit.

Initially, during the quiet corona interval (panel~a, 03:20--03:45~UT, identified by a smooth and nearly flat soft X-ray flux curve), the plasma exhibits typical coronal abundances, with measured FIP biases of Ca ($\sim$4.0), S ($\sim$1.3), and Ar ($\sim$1.8).
These values, enhanced relative to their photospheric abundances owing to the FIP effect, provide a basal condition of the quiet-Sun about 4~hr prior to the associated subflare. %
Afterwards, during the main phase of the subflare (panel~b, 07:27--07:33~UT), the bias of the low-FIP element Ca decreases slightly to $\sim$3.0.
Such a relative depletion of low-FIP elements in the corona is consistent with the process of chromospheric evaporation, as suggested in earlier studies~\citep[e.g.,][]{mondal2021EvolutionElementalAbundances, mithun2022SoftXRaySpectral, suarezEstimationsElementalAbundances2023, ngUnveilingMassTransfer2024}.
In the subsequent interval covering the decay to post-decay phase of the subflare (panel~c, 07:33--08:05~UT), the Ca bias rises markedly to $\sim$6.3, notably exceeding the typical coronal value and exhibiting an anomalously strong FIP effect.
However, it would be expected to recover toward its preflare coronal level (i.e., $\sim$4.0) if flare-induced processes are in charge of the variation (see references therein). %
This unexpected behavior arouses particular interest and motivates further investigation coordinated with EIS observations, which will be presented in the following section.

\subsection{Spatially Resolved Plasma Composition Diagnostics} \label{sec_eisFIP}

\begin{figure*}[ht!]%
\plotone{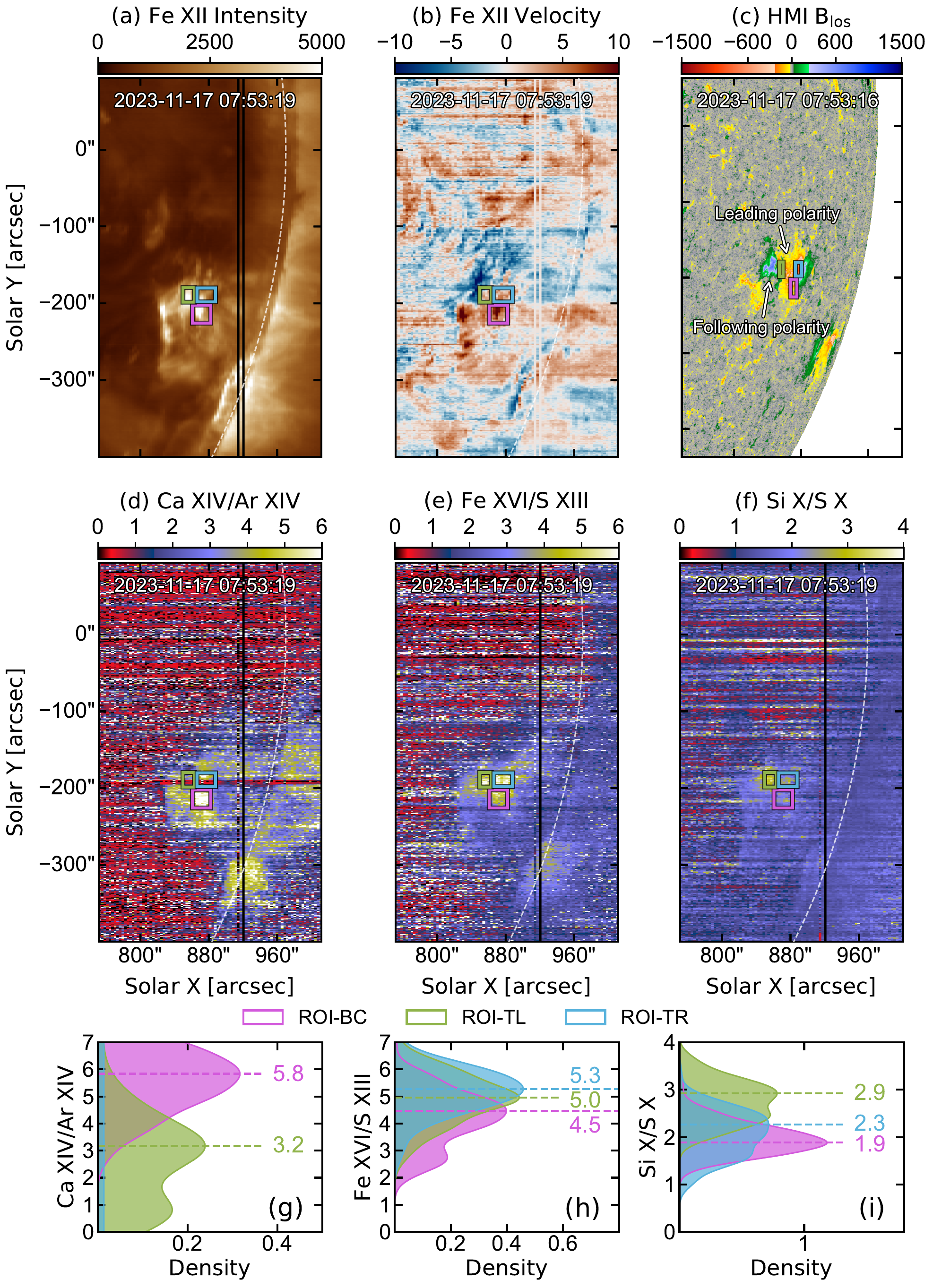}
\caption{Maps of (a) intensity and (b) Doppler velocity from \ion{Fe}{12} 195.12~{\AA} line, as measured by Hinode/EIS.
Velocities are given in units of $\mathrm{km~s^{-1}}$, with blue/red colors in panel~(b) indicating plasma motion (or emission) toward/away from the observer, respectively.
To provide context for the photospheric magnetic field, the LOS magnetogram from SDO/HMI, selected at the time closest to the start of the EIS raster observation, is shown in panel~(c).
The middle row displays FIP bias maps derived from the line ratios: (d) \ion{Ca}{14} 193.87~{\AA}/\ion{Ar}{14} 194.40~{\AA}, (e) \ion{Fe}{16} 262.98~\AA/\ion{S}{13} 256.69~{\AA}, and (f) \ion{Si}{10} 258.38~{\AA}/\ion{S}{10} 264.23~{\AA}.
The bottom row (g)--(i) shows the corresponding KDE distributions of pixel values within the ROIs: ROI-BC, ROI-TL, and ROI-TR, indicated by purple, green, and blue rectangles, respectively, on the FIP bias maps.
For each ROI, the peak KDE value is denoted by a dashed horizontal line with the value labeled adjacent to the line.
}{}
\label{fig_FIP_20231117_075319_KDE_overlaidBlos}
\end{figure*}

We present the observational results obtained from Hinode/EIS in Figure~\ref{fig_FIP_20231117_075319_KDE_overlaidBlos}, observed on 2023 November 17 during 07:53--08:54~UT, i.e., after the associated subflare in AR~13486.
The details of the raster scan are summarized in Table~\ref{tab_EIS_studies}.
Figure~\ref{fig_FIP_20231117_075319_KDE_overlaidBlos} consists of five diagnostic maps: the \ion{Fe}{12} intensity (panel~a), the \ion{Fe}{12} Doppler velocity (panel~b), and the elemental composition maps derived from the ratios \ion{Ca}{14}/\ion{Ar}{14} (panel~d), \ion{Fe}{16}/\ion{S}{13} (panel~e), and \ion{Si}{10}/\ion{S}{10} (panel~f), together with the corresponding kernel density estimation (KDE) distributions of the pixel values from the composition maps, shown in panels~(g)--(i).
These ratios serve as diagnostics of the FIP bias, reflecting variations in elemental abundances associated with the transport of mass and energy between the corona and the lower solar atmosphere.
For the context, panel~(c) shows the line-of-sight (LOS) magnetogram observed by Helioseismic and Magnetic Imager~\citep[HMI;][]{scherrerHelioseismicMagneticImager2012, schouDesignGroundCalibration2012} onboard the SDO, acquired at the beginning of the raster scan, with annotations indicating the leading and following magnetic polarities.
In the composition maps (panels~(d)--(f)), we identified three particularly distinct regions and defined them as regions of interest (ROIs), namely, ROI-BC (i.e., bottom center), ROI-TL (i.e., top left), and ROI-TR (i.e., top right), outlined by the purple, green, and blue rectangles, respectively.
These ROIs also exhibit bright low-FIP \ion{Fe}{12} line emission, as seen in Figure~\ref{fig_FIP_20231117_075319_KDE_overlaidBlos}(a).
In addition, when combined with AIA and HMI observations (see Figures~\ref{fig_event} and \ref{fig_FIP_20231117_075319_KDE_overlaidBlos}(c)), ROI-BC (between {\textminus 350} and 30~G) is situated within a region of relatively weak negative-polarity (leading) magnetic fields, corresponding to solar plage that exhibits brightening in the AIA 1600~{\AA} channel.
By contrast, ROI-TR ({\textminus 250} to {\textminus 900}~G) coincides with the sunspot umbra, characterized by strong negative-polarity (leading) fields and reduced emission in the 1600~{\AA} channel.
Moreover, ROI-TL (between {\textminus 430} and 30~G) exhibits properties intermediate between those of ROI-BC and ROI-TR, reflecting a transitional state between plage and sunspot environments.

To quantify the statistical properties of the elemental composition in these regions, the kernel density estimation distributions of the pixel values were further constructed from the composition maps. %
The peak values of these distributions are particularly noteworthy, since such measured abundances are higher than typical coronal levels. Specifically, ROI-BC exhibits a Ca/Ar ratio of approximately 6, ROI-TL a ratio of about 3, and all three ROIs show Fe/S ratios close to 5, whereas the Si/S ratios remain within the typical coronal range of 2--3. %
These diagnostics therefore indicate a pronounced enhancement of the FIP bias in the Ca/Ar and Fe/S ratios, but not in the Si/S ratio.
The EIS observations are broadly consistent with the MSS-1B Sun-as-a-Star measurements (see Figure~\ref{fig_mssSpec20231117}).
However, direct comparison requires caution, as the EIS elemental ratios represent relative abundances, i.e., the ratios of low- to mid/high-FIP elements, whereas the FIP bias values measured by MSS-1B represent absolute abundances, i.e., the abundances of elements expressed relative to hydrogen.
Notably, the EIS raster began after the associated subflare and covered a long time span, therefore the results collectively suggest that the anomalously strong FIP effect persisted for at least tens of minutes following the event. %

\subsection{Wavelet Analysis} \label{sec_wavelet}

\begin{figure*}[ht!]%
\plotone{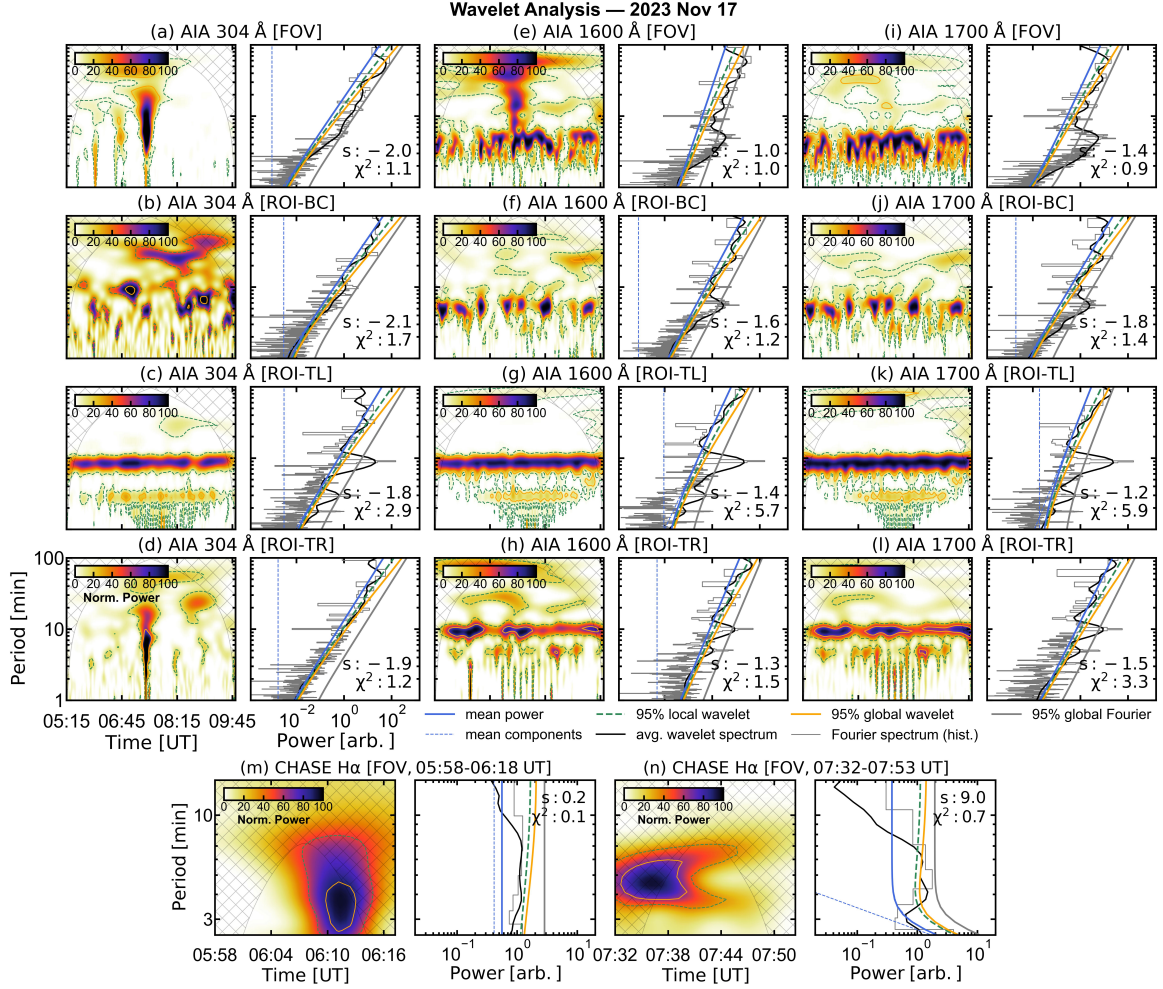}
\caption{Periodicity of AIA and CHASE light curves extracted from selected ROIs. %
Each column corresponds to the same region of interest (FOV, defined in Figure~\ref{fig_event}; ROI-BC, ROI-TL, and ROI-TR, as in Figure~\ref{fig_FIP_20231117_075319_KDE_overlaidBlos}), while each row corresponds to the same wavelength: AIA 304~{\AA} (a--d), 1600~{\AA} (e--h), and 1700~{\AA} (i--l), over the observation period from 05:15 to 09:50~UT on 2023 November 17.
The two bottom panels show CHASE H$\alpha$ line-core observations in the FOV defined in Figure~\ref{fig_event} from 05:58--06:18~UT (left) and 07:32--07:53~UT (right) on 2023 November 17.
In each panel, the left column shows the Morlet wavelet power spectrum, whitened by the background model and scaled for display.
The green dashed and yellow solid contours enclose regions where the wavelet power exceeds the 95\% local and global confidence levels, respectively.
The gray hatched region denotes the cone of influence (COI), where edge effects dominate and the period signals are not statistically significant.
The right column presents the time-averaged wavelet (black curve) and Fourier (gray histogram) power spectra.
The background model fitted to the Fourier spectrum is shown by the blue solid line, with its mean components indicated by the blue dashed lines.
The corresponding fitted power-law index $s$ and the reduced chi-square ($\chi^2$) value are displayed in the bottom-right (or top-right) corner.
The 95\% local and global time-averaged confidence levels are indicated by the green dashed and yellow solid lines, respectively, while the 95\% confidence level of the Fourier spectrum is shown by the gray solid line.
}{}
\label{fig_AIA_HA_wavelet}
\end{figure*}

Figure~\ref{fig_AIA_HA_wavelet} presents the periodicities identified by the wavelet analysis of the light-curve time series from the AIA 304~{\AA}, 1600~{\AA}, and 1700~{\AA} channels, spanning 05:15 to 09:50~UT on 2023 November 17, as well as the CHASE H$\alpha$ line-core observations.
The light curves are extracted from the FOV ([750\arcsec, \textminus280\arcsec] to [1000\arcsec, \textminus70\arcsec] in the helioprojective coordinates, as defined in Figure~\ref{fig_event}) and from the three ROIs defined in Figure~\ref{fig_FIP_20231117_075319_KDE_overlaidBlos}, namely, ROI-BC, ROI-TL, and ROI-TR.
Due to a data gap in the CHASE H$\alpha$ line-core observations, the analysis is carried out separately for 05:58--06:18~UT and 07:32--07:53~UT. For reliability, only the light curves from the FOV are included in the present study.
Of note here is that the CHASE H$\alpha$ line center forms in the chromosphere at an altitude of $\sim$1340~km above the photosphere~\citep{raoHeightdependentDifferentialRotation2024}, while the formation heights of AIA UV 1600 and 1700~{\AA} channels range from about 247 to 468~km above the photosphere~\citep{sanjay2024FormationHeightLowcorona}.
It is revealed that, in the 304~{\AA} channel (panels~(a)--(d)), the interval of strong wavelet power in both the FOV and ROI-TR coincide with the timing of the associated subflare, suggesting that these signals correspond to the flare-related pulsations. In contrast, the absence of such pulsations in ROI-BC and ROI-TL indicates that their own oscillatory signatures are more likely attributable to wave activity in the corona rather than direct flare emission.
In the 1600~{\AA} channel, a similar flare-related pulsation signature is detected in the FOV but is weak or absent in the three ROIs, indicating that the observed signals primarily reflect photospheric oscillations, without significant contamination from flare-related coronal emission.
In particular, as shown in panels~(f)--(h), ROI-BC exhibits quasi-periodic modulation, with the dominant power of 5-min oscillations in the photosphere, strongly suggestive of localized wave activity (e.g., leakage of global $p$-mode waves), whereas ROI-TL and ROI-TR display significant periodicity that remains nearly constant over time. %
The 1700~{\AA} channel, moreover, shows a broadly similar wavelet pattern, with oscillatory signals present but without a pronounced flare-related pulsation. This confirms that the detected signals from the respective regions can be used exclusively as diagnostics of photospheric oscillatory behavior.
Furthermore, as illustrated in panels~(m)--(n), the CHASE H$\alpha$ line-core observations reveal that the significant periodicities change markedly from 3.5~min (05:58--06:18~UT) before the associated subflare to $\sim$5~min (07:32--07:53~UT) after the flare, indicating chromospheric responses to the subflaring activity, which will be discussed in Section~\ref{sec_inclined}.

We further quantified the dominant periods through averaging of all pixels enclosed within the 95\% global confidence contours of the time-period (time-frequency) map (Figure~\ref{fig_AIA_HA_wavelet}).
For each isolated significant region, the mean period was calculated with pixels weighted by their wavelet power, while the spread was estimated as the standard deviation under a Gaussian assumption.
When multiple isolated significant regions were identified, their mean periods were combined into a single representative value using an inverse-variance weighted average.
Note that the significant regions taken into account are restricted to $1.5 < P < 10$~min, which corresponds to our study range of interest.
The averaged results of the wavelet analysis are summarized in Table~\ref{tab_wavelet}. %

\begin{deluxetable}{ccccc}[t]
\tabletypesize{\footnotesize}
\centering
\tablecolumns{5}
\tablecaption{Results of the wavelet analysis.
Mean periods (min) with $1\sigma$ standard deviations, estimated under a Gaussian assumption, are listed for each region of interest and wavelength, corresponding to Figure~\ref{fig_AIA_HA_wavelet}. \label{tab_wavelet}}
\tablehead{
\nocolhead{Wavelength} & 
\colhead{304~{\AA}} & 
\colhead{1600~{\AA}} & 
\colhead{1700~{\AA}} & 
\colhead{H$\alpha$ line-core} \\
\cline{2-5}
\colhead{Region} & 
\multicolumn{4}{c}{Mean Period (min)}
}
\startdata
\\[-8pt]
FOV    & 7.73 (0.59) & 4.44 (0.03) & 3.23 (0.02) & \makecell[l]{3.50 (0.06)\tablenotemark{a}\\4.69 (0.07)\tablenotemark{b}} \\
ROI-BC & 7.52 (0.06) & 5.12 (0.02) & 5.11 (0.02) & \nodata \\
ROI-TL & 8.79 (0.07) & 8.60 (0.11) & 8.58 (0.17) & \nodata \\
ROI-TR & 4.40 (0.33) & 9.77 (0.06) & 9.57 (0.07) & \nodata \\
\enddata
\tablenotetext{a}{Observed from 05:58 to 06:18~UT; see Figure~\ref{fig_AIA_HA_wavelet}(m).}
\tablenotetext{b}{Observed from 07:32 to 07:53~UT; see Figure~\ref{fig_AIA_HA_wavelet}(n).}
\end{deluxetable}

\section{Discussion and Physical Interpretation} \label{sec_discussion}

\subsection{Reconfigured Local Inclined Fields} \label{sec_inclined} %

In Table~\ref{tab_wavelet}, our wavelet analysis of the CHASE H$\alpha$ line-core observations reveals that the chromospheric mean wave period within the AR increased from 3.5~min prior to the associated subflare (05:58--06:18~UT) to $\sim$5~min after the subflare (07:32--07:53~UT).
We conjecture that the increase of the oscillation period is due to a reconfiguration of the chromospheric magnetic field structure following the flare.
The increase in the dominant wave period suggests that the local magnetic field became more inclined (i.e., less perpendicular to the solar surface), because a more inclined field lowers the cutoff frequency for $p$-mode waves, allowing longer-period waves to propagate more easily~\citep{roberts2006SlowMagnetohydrodynamicWaves}.
One caveat is the low temporal cadence of CHASE observations ($\sim$1~min), which may be only marginally sufficient for interpreting the data in support of this conjecture.

\begin{figure*}[ht!]%
\plotone{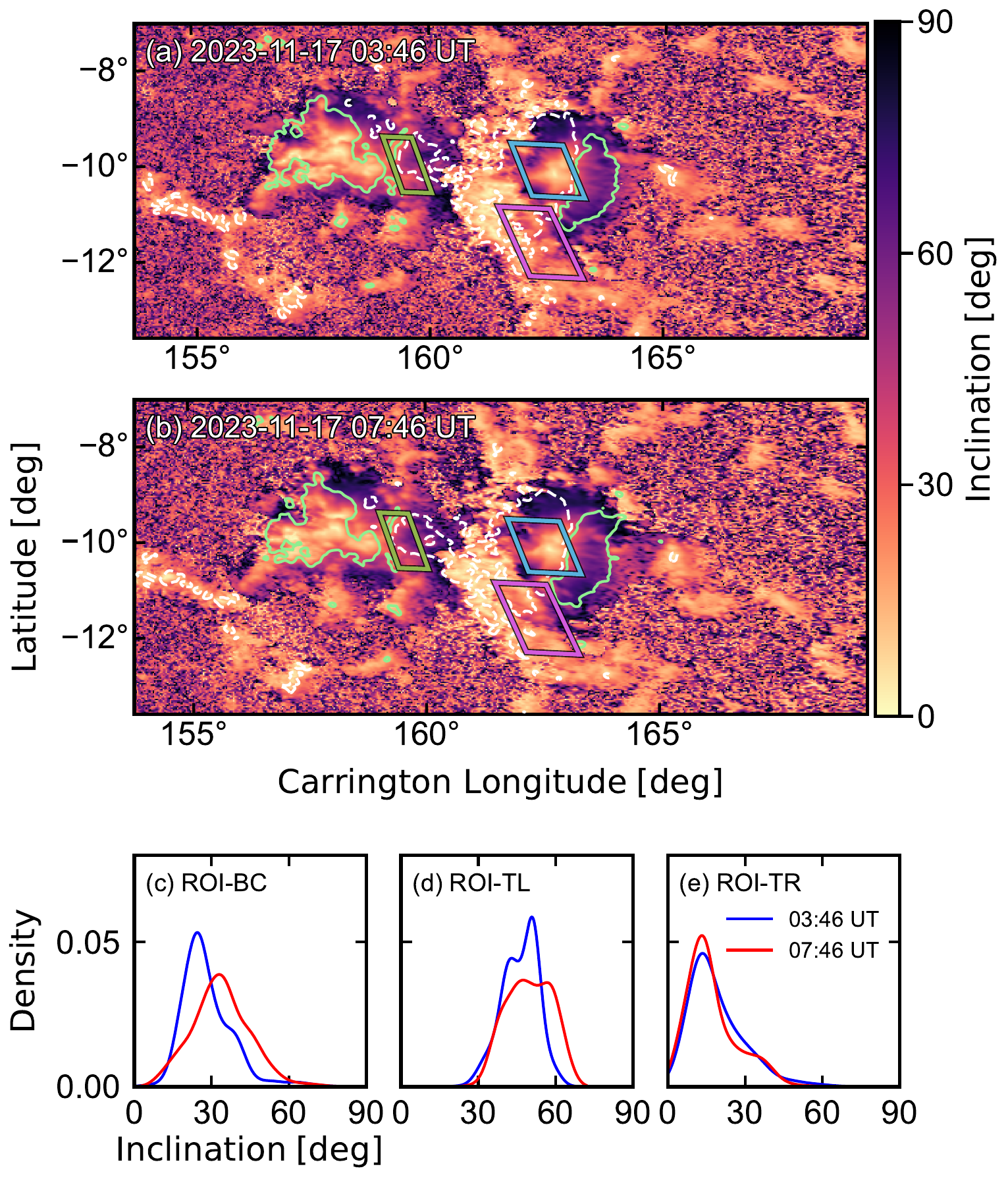}
\caption{Inclination angle distributions derived from SDO/HMI SHARP vector magnetograms for (a) 03:46~UT and (b) 07:46~UT on 2023 November 17.
The three ROIs---ROI-BC, ROI-TL, and ROI-TR---are outlined by purple, green, and blue polygons, respectively.
The overlaid contours show the LOS magnetic field at $\pm150$~G, with green and white indicating positive and negative polarities, respectively.
The bottom panels show the corresponding KDE distributions for (c) ROI-BC, (d) ROI-TL, and (e) ROI-TR, with the 03:46~UT and 07:46~UT intervals denoted as the blue and red curves, respectively.
}
\label{fig_binc_stats}
\end{figure*}

To confirm our conjecture, we utilized the Space-weather HMI Active Region Patches~\citep[SHARP;][]{bobraHelioseismicMagneticImager2014} vector magnetograms from the Cylindrical Equal-Area (CEA) data series (\texttt{hmi.sharp\_cea\_720s}) to determine the evolution of the magnetic field inclination, i.e., the angle between the vector magnetic field and the local radial direction.
Figure~\ref{fig_binc_stats} reveals the inclination angle maps for 03:46~UT (preflare; top panel) and 07:46~UT (postflare; middle panel), corresponding to the fitted intervals used in the MSS-1B spectral analysis (see Figure~\ref{fig_mssSpec20231117}).
To quantify these changes, we analyzed the inclination statistics within the three ROIs using kernel density estimation, weighting each pixel by its distance from the ROI center.
The resulting distributions are presented in the bottom panel of Figure~\ref{fig_binc_stats}, with the preflare interval (03:46~UT) in blue and the postflare interval (07:46~UT) in red.
Among the three ROIs, the most prominent change occurs in ROI-BC (plage region; panel~c), where the peak inclination increases from approximately 25$^\circ$ in the preflare interval to about 33$^\circ$ in the postflare interval.
In contrast, ROI-TR (sunspot umbra; panel~e) maintains a nearly invariant inclination distribution, with a maximum inclination of about 13$^\circ$ to the normal (i.e., close to vertical relative to the solar surface).
Meanwhile, the inclinations in ROI-TL persistently fall between 30$^\circ$ and 60$^\circ$, with their kernel density estimation skewed toward more inclined orientations.
Thus, the photospheric magnetic measurements shown in Figure~\ref{fig_binc_stats}, combined with the chromospheric diagnostics summarized in Table~\ref{tab_wavelet}, provide compelling evidence that the associated subflare led to an increase in the field inclination after magnetic reconnection. %

\begin{figure}[ht!]%
\plotone{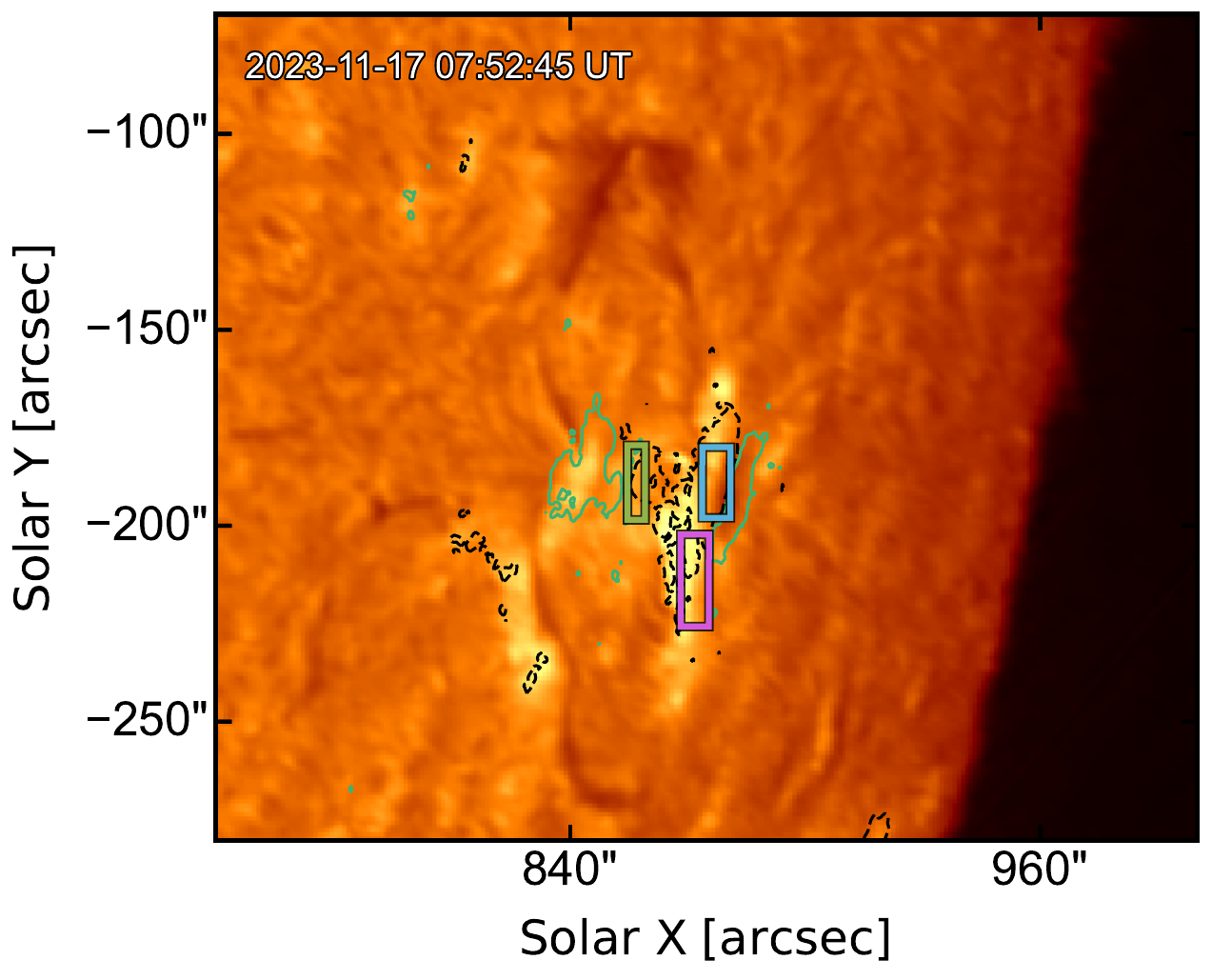}
\caption{CHASE/HIS 6562.8~{\AA} map taken at 07:52~UT.
Contours of the LOS magnetic field at $\pm150$~G are overplotted, with green and black colors denoting positive and negative polarities, respectively.
The purple, green, and blue rectangles denote the ROIs: ROI-BC, ROI-TL, and ROI-TR, respectively.
}
\label{chaseHA_20231117_075245_update}
\end{figure}

\subsection{What Drives the Anomalously Strong FIP Effect?}

On the basis of the above results, the observed photospheric oscillations with a dominant 5-min periodicity (Table~\ref{tab_wavelet}) are plausibly interpreted as the global $p$-mode waves that leak upward along the inclined magnetic structure. %
The features reported in the present study are similar to those identified by \citet{depontieuSolarChromosphericSpicules2004} and \citet{wangPROPAGATINGSLOWMAGNETOACOUSTIC2009}, supporting a common origin in the upward leakage of $p$-mode waves.
As further noted by \citet{depontieuSolarChromosphericSpicules2004}, the inclined geometry of the flux tubes is particularly crucial, as it significantly enhances the efficiency with which the long-period $p$-mode waves can tunnel upward from the photosphere. %
At the equipartition layer, where the sound speed ($v_s$) and Alfv\'en speed ($v_A$) become comparable, mode conversion occurs due to the coupling of restoring forces.
This process converts a fraction of the slow-mode waves (i.e., upward-leaking $p$-mode waves) into fast-mode waves (e.g.,~\citealp{callyDispersionRelationsRays2005}; \citealp{schunkerMagneticFieldInclination2006}; \citealp{callyThreeDimensionalMHDWave2008}; \citealp{callyBENCHMARKINGFASTTOALFVENMODE2011}; \citealp{khomenkoNUMERICALSIMULATIONSCONVERSION2012}; see also the reviews by \citealp{srivastavaChromosphericHeatingMagnetohydrodynamic2021} and \citealp{mortonAlfvenicWavesInhomogeneous2023}).
The remainder continues to propagate upward as slow modes guided along the magnetic field lines.
Subsequently, these guided wave modes are efficiently transmitted upward into the chromosphere and continue their upward propagation.
While a portion of the wave flux reflects from regions with steep density gradients, the remainder couples into the coronal loop through the footpoints.
Once inside the coronal loop, the waves travel along magnetic field lines, reflecting at the upper chromosphere at the opposite footpoint and bouncing back and forth within the loop structure. For the long-period waves to resonate with the coronal loops, the magnetic loops should be long, which is verified with the extrapolated potential field distribution.
This process gives rise to an upward-directed ponderomotive force that separates chromospheric low-FIP ions from high-FIP neutrals and lifts them into the corona, resulting in the observed FIP effect.

For the FIP effect to be anomalously strong, either the upward transport of low-FIP ions into the corona or the elemental fractionation occurring in the lower solar atmosphere must be exceptionally efficient.
Based on the analysis of the observational data, we suggest the following causal chain for the anomalously strong FIP effect observed.
The principal impediment to the fractionation process is the plasma-$\beta$$\simeq$1 layer, which can partially reflect upward-propagating waves and generate a competing downward-directed ponderomotive force. Note that the plasma-$\beta$=$2/\gamma$ corresponds to $v_A$=$v_s$, where $\gamma$=$5/3$ is the adiabatic index (i.e., plasma-$\beta$=1.2).
We propose that the magnetic field lines become more inclined after magnetic reconnection related to the subflare.
We regard the anomalously strong FIP effect observed in ROI-BC (following the subflaring episode) as compelling evidence that this magnetic channel is optimized for efficient wave transmission. %
The inclined magnetic geometry appears to play a dual role: it not only allows more $p$-mode waves with longer periods to leak from the photosphere to the chromosphere by lowering the acoustic cutoff frequency but also configures the chromosphere in a way that increases upward-propagating MHD wave transmission at the plasma-$\beta$$\simeq$1 layer, or say, reduces reflection at the  plasma-$\beta$$\simeq$1 layer.
As a result, these long-period MHD waves, channeled upward into the coronal loop, are repeatedly reflected at the upper chromosphere at the opposite footpoint.
This leads to the consequential net ponderomotive force that is strongly directed upward in a time-averaged sense, lifting the low-FIP ions into the corona and producing the highly fractionated (low-FIP--enriched) plasma, thereby giving rise to the observed anomalously strong FIP effect in the corona.
Moreover, the height of the plasma-$\beta$$\simeq$1 layer may give an additional pathway for enhanced wave transmission if $p$-mode waves can mode convert to fast-mode waves before arriving at the acoustic cutoff region, typically located near the temperature minimum at the bottom of the chromosphere.
The reduction in the cutoff frequency due to magnetic inclination increases the height of wave propagation, extending to $\sim$430~km above the photosphere, as observed in plages and sunspots by \citet{rajaguru2019MagneticFieldsSupply}.
Given the magnetic field strength in the plage region (Figure~\ref{fig_FIP_20231117_075319_KDE_overlaidBlos}(c)), the plasma-$\beta$$\simeq$1 layer is likely located below, or at a similar height to, the acoustic cutoff region.
If this is the case, the initially acoustic waves spend a longer time near the plasma-$\beta$$\simeq$1 layer, and because acoustic waves in the photosphere and low chromosphere generally propagate radially, inclined magnetic fields result in a larger perpendicular wavevector component ($k_\perp$).
Both effects therefore enhance mode conversion in this layer for more inclined magnetic fields, which may contribute to a stronger FIP effect, as similarly discussed in \citet{lamingElementAbundancesImpulsive2023}.

In addition, the persistence of the anomalously strong FIP effect for tens of minutes is a key issue of interest and is likely associated with continued heating in the chromosphere. %
Prior studies have indicated that an inclined magnetic field in the photosphere is conducive to strong chromospheric heating over plage regions through wave-related processes such as mode conversion and dissipation~\citep{jefferiesMagnetoacousticPortalsBasal2006, stangaliniMHDWaveTransmission2011, khomenkoThreedimensionalSimulationsSolar2018, ananMeasurementsPhotosphericChromospheric2021}.
Magnetic reconnection, including that associated with magnetic flux cancellation, has also been suggested as another important heating mechanism in the lower solar atmosphere~\citep{chen2001EllermanBombsType, shibataChromosphericAnemoneJets2007, isobe2008ConvectiondrivenEmergenceSmallScale, leake2013MagneticReconnectionWeakly, priestCancellationNanoflareModel2018, tang2025HighresolutionObservationsSmallscale, huangUnifiedModelSolar2025}. %
For a thorough overview of the various physical heating mechanisms involving chromospheric magnetic reconnection, MHD waves, and plasma instabilities, see the reviews by \citet{ni2020MagneticReconnectionPartially} and \citet{srivastavaChromosphericHeatingMagnetohydrodynamic2021}.
In ROI-BC, identified as a plage region (Figures~\ref{fig_event}(q)--(t) and \ref{fig_FIP_20231117_075319_KDE_overlaidBlos}(c)), significant brightening was observed in the chromospheric H$\alpha$ 6562.8~{\AA} line (Figure~\ref{chaseHA_20231117_075245_update}), indicating localized strong heating within the plage chromosphere.
Following the associated subflaring episode, we also clearly observed the continuous occurrence of small-scale brightenings in the AIA 1600 and 1700~{\AA} channels surrounding the sunspot periphery as well as in the adjacent plage region, accompanied by corresponding flux cancellation in the HMI SHARP LOS magnetograms within the AR, indicating ongoing magnetic flux cancellation between oppositely polarized elements that collide in the photosphere. %
Therefore, it is most likely that the heating results from energy released through magnetic reconnection that occurs in the lower atmosphere (e.g.,~\citealp{chen2001EllermanBombsType}, \citealp{priestCancellationNanoflareModel2018}, \citealp{tang2025HighresolutionObservationsSmallscale}, \citealp{huangUnifiedModelSolar2025} and references therein).
Furthermore, such chromospheric heating is typically observed to be accompanied by patches of intense brightenings in the H$\alpha$ and \ion{Mg}{2}~k lines (appearing either in the wings without significant features in the core, or in both core and wings), as reported in earlier works~\citep[e.g.,][]{fang2006SpectralAnalysisEllerman, tian2016AREIRISBOMBS, ananMeasurementsPhotosphericChromospheric2021}. %
Our H$\alpha$ wing (6562.8~{\AA}$\pm$0.5--1.0~{\AA}; not shown) observations near the sunspot periphery and within portions of the plage reveal (1) clear bright points prior to the subflare (05:58--06:18~UT), and (2) only comparatively weak wing brightenings during the main-to-decay phase of the subflare (07:32--07:53~UT). %
Taken together, these combined AIA and CHASE H$\alpha$ diagnostics suggest that the reconnection current sheet during the subflare is located in the mid (or near-upper) chromosphere, where the plasma is partially ionized. Reconnection at this height produces bidirectional outflows, where the upward component in conjunction with energy released at the current sheet, heats the upper chromosphere and results in localized strong H$\alpha$ core emission, while the downward outflow component compresses and weakly heats the lower chromosphere and upper photosphere, producing the observed bright points in the AIA 1600 and 1700~{\AA} channels and the comparatively weak H$\alpha$ wing response. %

Moreover, the height of chromospheric reconnection has also been examined by \citet{huangUnifiedModelSolar2025}, who performed two-dimensional MHD simulations and demonstrated that when the preexisting magnetic field in an ephemeral flux region is slightly more inclined, the reconnection site resides in the upper chromosphere, whereas a more vertical field leads to reconnection occurring lower in the chromosphere.
The inclination angle adopted in their simulations was approximately 30$^\circ$ (private communication).
This is consistent with our measurements, namely, the inclination of $\sim$33$^\circ$ in ROI-BC and the inclination range between 30$^\circ$ and 60$^\circ$ observed in ROI-TL (Section~\ref{sec_inclined}).
Therefore, our observations suggest that the small-scale chromospheric reconnection before the subflare likely occurs in the low (or mid) chromosphere, and that the subsequent magnetic reconfiguration triggered by the subflare elevates the reconnection site into the mid (or near-upper) chromosphere.
The quantitative agreement between their simulation results and our observationally derived field inclinations further strengthens our interpretation that the enhanced magnetic field inclination plays an important role in facilitating the anomalously strong FIP effect and in sustaining the reconnection-driven heating discussed above in this work.
We further take note here that an additional contributing factor may be that magnetic reconnection tends to occur more frequently in regions with inclined magnetic fields, consistent with our observation of the continuous sequence of small-scale brightenings.
In short, the sustained heating in the upper/mid chromosphere is a key factor in supplying low-FIP ions from the upper chromosphere into the corona, as also suggested in our previous works~\citep{ngUnveilingMassTransfer2024, ngPeculiarFeatureFirst2025}. Such prolonged heating enables the highly fractionated plasma to be continuously transported upward and to persist within the corona. %

Furthermore, the difference in the evolution of the magnetic inclination has important implications for the observed FIP fractionation, as reflected in the contrasting diagnostic Ca/Ar line ratios of $\sim$6 (unusually enhanced, coronal-level fractionation) in ROI-BC and $\sim$3 (typical coronal-like level) in ROI-TR.
In terms of the solar features cospatial with these regions, ROI-BC is located in a plage region and exhibits a maximum inclination of about 33$^\circ$ in the postflare interval (roughly 25$^\circ$ before the subflare), whereas ROI-TR lies within a sunspot umbra, where the field inclination remains narrowly peaked at about 13$^\circ$ (i.e., nearly perpendicular to the solar surface), as described in Sections~\ref{sec_eisFIP} and \ref{sec_inclined}.
This indicates that the subflare-related reconnection produces only minimal changes in the umbral field inclination.
Since sunspot umbrae are known to host strongly vertical magnetic fields, it is natural that the umbral field in ROI-TR acts to anchor and locally pull the field lines back toward a more vertical orientation, even though the surrounding field lines are more inclined (Figure~\ref{fig_binc_stats}).
Consequently, such an overall inclined geometry can facilitate the efficient upward transmission of MHD waves in ROI-BC, and thus enhance the ponderomotive acceleration of low-FIP ions, while the locally vertical umbral field in ROI-TR can inhibit wave transmission, thereby reducing the efficiency of positive FIP fractionation.

In addition, there might be other possibilities for the enhancement of the low-FIP Fe/S bias to $\sim$5 in ROI-TR and ROI-TL.
First, the footpoints of the subflare are located near these two ROIs, hence the subflare might generate MHD waves as evidenced in Figures~\ref{fig_AIA_HA_wavelet}(d), (h), and (i), which can push low-FIP elements into the corona. However, such MHD waves do not last long and therefore cannot explain why the FIP effect sustained for a long time well after the subflare.
Rather, it is speculated that the enhanced low-FIP Fe/S bias is more likely associated with the loop--loop interaction due to magnetic reconnection responsible for the subflare, which involves one of the pre-existing loops with a high Fe/S bias.
In the scenario proposed by \citet{toSpatiallyResolvedPlasma2024}, reconnection simultaneously creates a downward outflow of high Fe/S--biased plasma that is subsequently confined near the flare loop top by evaporation upflows from both footpoints.
Hence, if this is the case, the enhanced Fe/S--biased plasma is possibly trapped near the subflare loop apex, leading to enhanced FIP effect in the subflare.
It is noted in passing that, as illustrated in Figure~\ref{fig_FIP_20231117_075319_KDE_overlaidBlos}(b), the coronal redshifts evident in the three ROIs correspond to downward-moving plasma that may be associated with the aforementioned reconnection downward outflows.

\section{Summary} \label{sec_summary}

Using multiwavelength observations of AR~13486 from the MSS-1B/SXDU, Hinode/EIS, CHASE/HIS instruments, alongside SDO/AIA and HMI on 2023 November 17, we present the evolution of plasma composition in response to a subflare. %
We propose that some magnetic field lines become more inclined, and the inclined magnetic configuration in this region enhances the transmission of upward-propagating MHD waves, effectively reducing reflection near the plasma-$\beta$$\simeq$1 layer, thereby significantly enhancing FIP fractionation produced by the consequential net upward-directed ponderomotive force. %
Furthermore, sustained chromospheric heating, driven by chromospheric magnetic reconnection associated with flux cancellation and the inclined geometry, was crucial in maintaining the enhanced FIP effect for tens of minutes following the event, indicating that dynamic reconnection processes may play an important role in regulating coronal plasma composition.
The spatial disparity in fractionation levels between plage and umbral regions further reinforces the significance of inclined magnetic field orientation as a key determinant of FIP fractionation.

\begin{acknowledgments}
This research was supported by the Science and Technology Development Fund (FDCT) of Macau (grant Nos.~0008/2024/AKP, 0034/2024/AMJ, 0008/2024/ITP1, 002/2024/SKL, 0002/2025/AKP) and NSFC (12127901).
This work makes use of data from the MSS-1 mission, supported by China National Space Administration (CNSA) and Macao Foundation, and from the CHASE mission, supported by CNSA.
Hinode is a Japanese mission developed by ISAS/JAXA with NAOJ as domestic partner, and NASA and STFC (UK) as international partners. It is operated by these agencies in cooperation with ESA and NSC (Norway).
SDO data were obtained courtesy of NASA/SDO and the AIA and HMI science teams.
CHIANTI is a collaborative project involving George Mason University, the University of Michigan (USA), University of Cambridge (UK) and NASA Goddard Space Flight Center (USA).
We appreciate the anonymous reviewer for providing insightful suggestions and helpful comments.
M.-H.N. would also like to thank Shihao Rao at Nanjing University for his generous assistance in the processing of the CHASE fake Doppler maps, which are related to this study but are not presented in the final manuscript.
\end{acknowledgments}

\software{OSPEX~\citep{tolbertOSPEXObjectSpectral2020}, EISPAC~\citep{webergEISPACEISPython2023}, CHIANTI~\citep{dereCHIANTIAtomicDatabase1997, dere2023CHIANTIAtomicDatabase}, SSW~\citep{freelandDataAnalysisSolarSoft1998}, aiapy~\citep{barnesAiapyPythonPackage2020}, Astropy~(The Astropy Collaboration et al.~\citeyear{theastropycollaborationAstropyProjectSustaining2022}), SunPy~(The SunPy Community et al.~\citeyear{thesunpycommunitySunPyProjectInteroperable2023}), emcee~\citep{foreman-mackeyEmceeMCMCHammer2013}.}

\bibliographystyle{aasjournalv7}

\end{document}